\documentclass[vecphys]{svmult}
\usepackage{epsfig}
\usepackage{graphicx} %,psfig}
\usepackage{amsmath,amssymb,amsfonts,textcomp}

\newcommand{\avr}[1]{\ensuremath{\left\langle#1\right\rangle}}

\newcommand{\bra}[1]{\ensuremath{\left\langle#1\right|}}
\newcommand{\ket}[1]{\ensuremath{\left|#1\right\rangle}}

\newcommand{\db}[1]{\ensuremath{\mathrm{d}#1}}

\title{Path integral Monte Carlo simulation 
of charged particles in traps}

\titlerunning{PIMC simulation 
of charged particles in traps}

\author{Alexei Filinov \and Jens B\"oning \and Michael Bonitz}
\institute{Christian-Albrechts-Universit\"at zu Kiel, Institut
f\"ur Theoretische Physik und Astrophysik, Leibnizstrasse 15,
24098 Kiel, Germany}

\begin{document}

\maketitle

\section{Introduction}

This chapter is devoted to the computation of equilibrium (thermodynamic) properties of  quantum systems. In particular, we will be interested in the situation where the interaction between particles is so strong that it cannot be treated as a small perturbation. For weakly coupled systems many efficient theoretical and computational techniques do exist. However, for strongly interacting systems such as nonideal gases or plasmas, strongly correlated electrons and so on, perturbation methods fail and alternative approaches are needed. Among them, an extremely successful one is the Monte Carlo (MC) method which we are going to consider in this chapter.

\section{Idea of Path integral Monte Carlo}

If we perform classical simulations of a system in equilibrium, we usually need a Boltzmann-type probability distribution, $p_{B}\sim e^{-\beta U_N(R)}/Z$, ($\beta=1/k_B T$) and then the Monte Carlo method~\cite{other,numbook06} can be used to sample the particle coordinates $R=(\mathbf r_1, \mathbf r_2, \ldots, \mathbf r_N)$. Now the question arises what is the appropriate probability density in the quantum case. The answer is provided by the density operator ${\hat\rho}$. Consider a general expression for thermodynamic averages in statistical thermodynamics.
The $N-$particle density matrix $\rho(\beta)$ contains the complete information about the system with the observables given by ($Tr$ denotes the trace)
\begin{equation}
\langle\hat{O}\rangle(\beta)=\frac{{{\rm Tr}
\left[\hat{O}\,\hat{\rho}\right]}} {{{\rm
Tr}\left[\hat{\rho}\right]}} =\frac{\displaystyle\int
\db{R} \,\langle R |\hat{O}\,\hat{\rho}| R \rangle}
{\displaystyle\int
\db{R} \,\langle R |\hat{\rho}(\beta)| R \rangle}=\frac{\displaystyle\int
\db{R} \,\mathrm{d} R'\,\langle R|\hat{O}|R'\rangle
\,\langle R'|\hat{\rho}| R \rangle} {\displaystyle\int
\db{R} \,\langle R |\hat{\rho}| R \rangle}.
\label{pimc_mean}
\end{equation}
This expression is simplified if the physical observable $\hat O$ is diagonal in the coordinate representation, i.e.
$
\langle R'|\hat{O}\,\hat{\rho}(\beta)| R \rangle=
\langle R |\hat{O}\,\hat{\rho}(\beta)| R \rangle \,
\delta( R'-R).
$

As in the classical case we need to perform an integration over $3N$ (or more) degrees of freedom. The problem here, compared to the classical case, is that, in general, we do know an analytical expression for the $N$-particle density operator in the coordinate representation which can be substituted in Eq.~(\ref{pimc_mean}). This problem was first overcome by Feynman~\cite{Feynm}. The key idea is to express the unknown density operator,
\begin{equation}
\langle R |{\hat\rho}(\beta)| R' \rangle \quad \text{with} \quad \hat\rho=e^{-\beta \hat H},
\end{equation}
 at a given inverse temperature $\beta$ by its high-temperature asymptotic which is known analytically. But the price to pay is high: instead of an already complicated $3N-$dimensional integral, now it expands to much higher dimensions ($3N\cdot M$), where $M$ is an integer which in practice is chosen as $1 \leq M \leq 3000$.

{\em \textbf{A. Group property of DM}}\\
One simple and straightforward strategy is to use the group property of the density matrix. It allows to express the density matrix at low temperatures in terms of its values at higher temperature, i.e.
\begin{align}
%\begin{split}
\rho (R,R';\beta _{1}+\beta_{2}) &=
	\langle R|e^{-(\beta_{1}+\beta_{2})\hat{H}}|R'\rangle=
	\int\!\db{R_1}\,\bra{R}e^{-\beta_{1}\hat{H}}\ket{R_1}\bra{R_1}e^{-\beta_{2}\hat{H}}\ket{R'}\notag\\
	&=\int\!\db{R_1}\,\rho(R,R_{1};\beta_1)\rho(R_{1},R';\beta_2).
%\end{split}
\end{align}
%\begin{equation*}
%\beta_1, \beta_2 < (\beta_{1}+\beta_{2}) \quad \Rightarrow \quad T_{1},T_{2} > T=1/k_B(\beta_1+\beta_2).
%\end{equation*}
The next step is a generalization: use the group property $M$ times
\begin{equation}
\hat{\rho} = \exp[-\beta \hat{H}] = \exp[-\Delta\beta\hat{H}]\ldots\exp[-\Delta \beta \hat{H}], \quad\Delta \beta =\beta /M.
\label{conv1}
\end{equation}
This means that the density operator $\hat \rho$ is expressed as a product of $M$ new density operators, $e^{-\Delta \beta {\hat H}}$, each corresponding to an $M$ times higher temperature.

Finally, using Eq.~(\ref{conv1}) we can write\footnote{The total dimension of the integral, $(M-1)$ $3 N$, may be very large. The success of the method relies on highly efficient Monte Carlo integration.} the off-diagonal matrix element as (for any fixed end-points $R$ and $R'$)
\begin{multline}
\rho (R,R';\beta) = \int\!\db{R_1}\db{R_2}\ldots\db{R_{M-1}}\;\\
\rho(R,R_{1};\Delta\beta)\rho (R_{1,}R_{2};\Delta \beta)\ldots\rho(R_{M-1},R';\Delta\beta),
\label{conv2}
\end{multline}
where $M$ factors are connected by $M-1$ intermediate integrations.

{\em \textbf{B. High temperature approximation}}\\
The Eqs.~(\ref{conv1}) and~(\ref{conv2}) are exact at any finite $M$ as long as we use exact expressions for the high-temperature $N$-particle density matrices, $\rho(R_{i-1},R_i;\Delta\beta)$. Unfortunatly, they are unknown, and to proceed further we need to introduce approximations.

Our approximation will be based on Trotter{\textquotesingle}s theorem (1959) applied to a general Hamiltonian, $\hat H =\hat T + \hat V$, which contains both kinetic and potential energy operators, i.e.
\begin{gather}
\hat{\rho } = e^{-\beta (\hat{T}+\hat{V})},
	\qquad\hat{\rho} = \lim_{M\rightarrow\infty}\left[e^{-\Delta\beta\hat{T}}e^{-\Delta\beta\hat{V}}\right]^{M} \\
\hat{\rho } \approx
	\left[e^{-\Delta \beta \hat{T}}e^{-\Delta \beta\hat{V}}\right]^{M}+O\left(e^{-\Delta \beta ^{2}M[\hat{T},\hat{V}]/2}\right)\approx
	\left[e^{-\Delta \beta \hat{T}}e^{-\Delta \beta \hat{V}}\right]^{M}+O(1/M).
\label{trotter}
\end{gather}
Note that $\hat T$ and $\hat V$ do not commute giving rise to the commutator, $[\hat{T},\hat{V}]$, which is only the first term of a series\footnote{Double, triple and high-order commutators in a series have as a prefactor higher powers $\Delta \beta^n$ and they can be dropped in the limit $\Delta \beta \rightarrow 0$.}. Neglecting  the terms $[{\hat T},{\hat V}]$ gives an error of the order $O\left[1/M\right]$. This error can be made arbitrarily small by choosing a sufficiently large number of factors $M$.

Using the Trotter result~(\ref{trotter}), we immediately obtain an approximation for high temperatures\footnote{Other more accurate high temperature approximations are discussed in Ref.~\cite{Cep95,numbook06}.}
\begin{align}
\rho (R_{i},R_{i+1};\Delta \beta ) &\approx
	\langle R_{i}|e^{-\Delta \beta\hat{T}}e^{-\Delta \beta \hat{V}}|R_{i+1}\rangle \notag \\
&= \lambda_{\Delta}^{-3N}\exp\left[-\frac{\pi}{\lambda _{\Delta}^{2}}(R_{i}-R_{i+1})^{2}-\Delta \beta V(R_{i};\Delta\beta)\right],
\label{coord_pr}
\end{align}
where $\lambda_{\Delta}=(2\pi\hbar^2 \Delta\beta/m)^{\frac 1 2}$ is the De~Broglie wavelength.
Substituting Eq.~(\ref{coord_pr}) in Eq.~(\ref{conv2}) we get our final result for low-temperatures
\begin{equation}
\rho (R,R';\beta) = \int\!\db{R_1}\ldots\db{R_{M-1}}\;
e^{-\sum\limits_{i=0}^{M-1}\frac{\pi}{\lambda _{\Delta}^{2}}(R_{i}-R_{i+1})^{2}}
e^{-\sum\limits_{i=0}^{M-1}\Delta \beta V(R_{i})},
\label{final}
\end{equation}
with the boundary conditions: $R_0=R$ and $R_{M}=R'$. Hence, we have succeeded in the analytical representation of the $N$-particle density matrix. This representation is then used in the numerical evaluation of this integral with  the help of the Monte Carlo based algorithm.

\begin{figure}[t]
\vspace{-1.3cm}
 \centering
 \includegraphics[height=10cm,angle=-90]{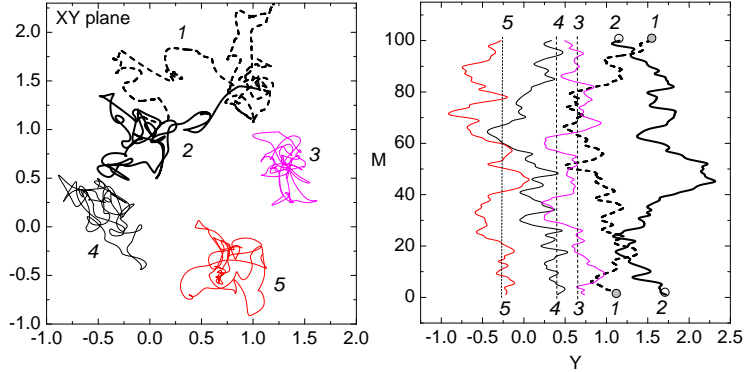}
 \vspace{-0.60cm} \caption{Live picture from a computer simulation. Five
 particles in a 2D parabolic potential. Each particle is presented as a continuous path (obtained by a smooth interpolation through a discrete set of $M=100$ points).
Right fig. shows how the paths are stretched along the time ($\beta$) axis. Particles 1 and 2 are in a pair exchange. Labels show particle
 indices.} \label{2d_5ferm}
 \end{figure}

{\em \textbf{C. Visualization of diagonal elements of the density matrix}}\\
As we can see from Eqs.~(\ref{conv2}) and~(\ref{final}), now all $N$ particles have their own images on $M$ different planes (or ``time slices''). We can view these images (for each particle $3\cdot M$ sets of coordinates) as a ``trajectory'' or a ``path'' in the configurational space. The inverse temperature argument $\beta$ can be considered as an imaginary time of the path. The set of $M$ time slices is ordered along the $\beta$-axis and separated by intervals $\Delta \beta$. On Fig.~\ref{2d_5ferm} we show typical configurations of particle trajectories which contribute to the diagonal density matrix element, Eq.~(\ref{conv2}) with $R=R'$. The full DM $\rho(R,R;\beta)$ is obtained after integration over all possible path configurations with the fixed end points $(R=R')$.

If we look at the final analytical result for the high temperature density matrix, Eq.~(\ref{final}), we recognize the usual Boltzmann factor with some effective action in the exponent. This action describes two types of interaction. The first term,
\begin{equation}
\sum_{i=0}^{M-1} \frac{\pi}{\lambda^2_{\Delta}}(R_{i}-
R_{i+1})^2=\frac{\pi}{\lambda^2_{\Delta}} \sum_{j=1}^{N}
\sum_{i=0}^{M-1} (\mathbf r_{i}^j- \mathbf r_{i+1}^j)^2 \label{kin_e_chang}
\end{equation}
comes from the kinetic energy density matrices of free particles ($j$ denotes the particle index). This energy can be interpreted as the energy of a spring, $U_s=\frac{k}{2}(\Delta x)^2$. Changing one of the coordinates is equivalent to a change of the spring energy. These springs provide that the nearest points on the path are usually at some average distance proportional to $\lambda_{\Delta}$. With temperature reduction the average size of the path increases with $\lambda_{\Delta}$.

The second term $\Delta \beta V(R_{i})$ in Eq.~(\ref{final}) adds interactions to the system (e.g. an external potential or inter-particle pair interaction)
\begin{equation}
\sum_{i=0}^{M-1} \Delta \beta V(R_{i})
= \Delta \beta \left( \sum_{j=1}^{N} \sum_{i=0}^{M-1} V_{ext}(\mathbf r_{i}^j)
+ \sum_{j < k} \sum_{i=0}^{M-1} V_{pair}(\mathbf r_{i}^j,\mathbf r_{i}^k)\right). \label{pot_e_chang}
\end{equation}
Each potential term depends only on the particle coordinates on the same time slice, i.e. $(\mathbf r_{i}^1,\mathbf r_{i}^2,\ldots,\mathbf r_{i}^N)$. As a result the number of pair interactions on each time slice, $N(N-1)/2$, is conserved.

In all expressions above we have considered particles as distinguishable ones. Generalization to quantum particles obeying Bose statistics is considered in Sec.~\ref{bosons}. The case of Fermi statistics is discussed e.g. in Ref.~\cite{Kleinert,fixed3,Cep95,numbook06}.

\section{Basic numerical issues of PIMC}
\label{issues}

Having the general idea of the PIMC simulations we are ready to formulate the first list of important issues which we need to solve:

{\em \textbf{A. How to sample the paths}}\\
It is necessary to explore the whole coordinate space for each intermediate point. This is very time consuming. To speed up convergence: move several slices (points of path) at once.

\begin{figure}[t] \vspace{-1.3cm}
 %\hspace{-1.0cm}
 \centering
 \includegraphics[height=12cm, angle=-90]{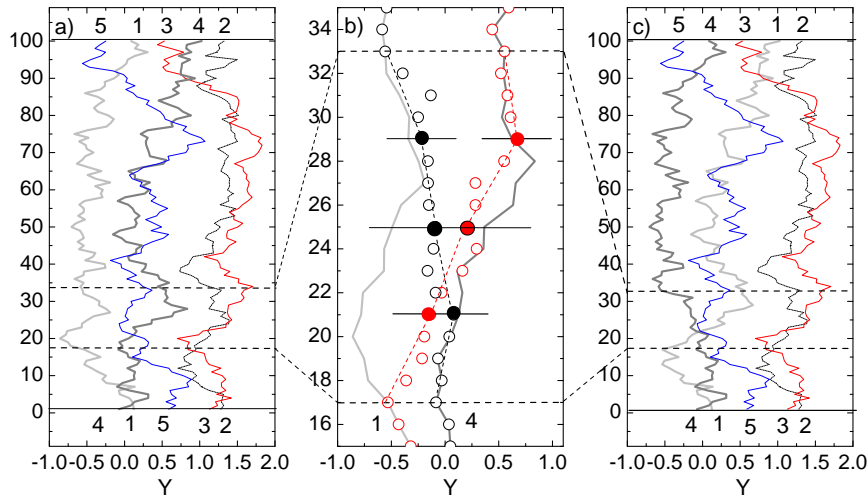}
\vspace{-0.4cm}
 \caption{(a),(c) The Y-coordinates of five identical particles as a function
 of the time-slice number $m$. Labels show particle indices. Thick
 gray and light gray lines show the paths of the particles ``$1$''
 and ``$4$'' which are exchanged by sampling new paths at
 time-slices $m=17-33$ (these time-slices are in the region between
 two dashed lines). (b) Sampling of new paths using the bisection
 algorithm for the electrons ``$1$'' and ``$4$''. The new paths are
 constrained at the time-slices $m=17-33$. Old (new) paths are
 shown by lines (circles). The filled circles show two mid-points
 sampled at the level $l=1$ (center of the interval, $m=25$) and
 four other mid-points for sub-intervals $[17,25]$ and $[25,33]$
 sampled at level $l=2$. Open circles show final new paths for two
 particles obtained with the sampling at levels $l=3,4$ and the
 transposition, i.e. by exchanging the paths starting from $m=33$
 up to the end point, $m=100$.} \vspace{-0.0cm} \label{bisec_exam}
 \end{figure}

Key point: sample a path using mid-points $R_{m}$ and a consequent iteration (\emph{bisection}), see Fig.~\ref{bisec_exam}(b). 

Using the definition: $0< t <\beta$, $\tau=i_0\Delta \beta\, [i_0=1,2,3,\ldots], R\equiv R(t), R' \equiv R(t+2\tau), R_{m}\equiv R(t+\tau)$; Guiding rule to sample a mid-point $R_{m}$ is
\begin{equation}
P(R_{m})=\frac{\langle R \vert e^{-\tau \hat H} \vert R_{m} \rangle \langle R_{m} \vert e^{-\tau \hat H} \vert R' \rangle}{\langle R \vert e^{-2 \tau \hat H} \vert R' \rangle}\approx (2\pi \sigma_{\tau}^2)^{-d/2} e^{-(R_{m}-\overline{R})^2/2 \sigma_{\tau}^2 }.
\label{samp}
\end{equation}
In practice, we can neglect in the sampling distribution the potential energy and use only the ratio of the free-particle density matrices. As a result we get a Gaussian distribution with the mean, $\overline{R}=(R+R')/2$ and the variance, $\sigma_{\tau}^2=\hbar^2 \tau/2m$. This will lead to $100 \%$ acceptance of sampling for ideal systems (and close to one for a weakly interacting system).

For strongly interacting systems the overlap of the paths sampled from the free-particle distribution~(\ref{samp}) results in large increase of the interaction energy and in a poor acceptance probability at the last level of the bisection sampling~\cite{Cep95,numbook06}. This can be improved by using the optimized mean and the variance
\begin{equation}
\overline{R}=(R+R')/2+ \sigma_{\tau} \frac{\partial V(\overline{R})}{\partial R}, \quad \sigma_{\tau}^2=\hbar^2 \tau/2m + \left(\frac{\hbar^2 \tau}{m} \right)^2 \Delta V(\overline{R}),
\end{equation}
which account for local interaction strength (nearest neighbor interaction).

Advantages of the bisection sampling method~\cite{Cep95,numbook06}:
\begin{itemize}
\setlength{\itemsep}{0.1mm plus3mm minus1mm}
\item Detailed balance property is satisfied at each level.

\item We do not waste time on moves for which paths come close and the potential energy strongly increases (for repulsive interaction). Such configurations are rejected already on early steps.

\item Computer time is spent more efficiently because we consider mainly configurations where the acceptance rate is high.

\item Sampling of particle permutations is easy to perform. 
\end{itemize}

{\em \textbf{B. Choose action as accurate as possible}}\\
It is better to use effective interaction potentials which take into account two, three and higher order correlation effects. More accurate actions help to reduce the number of time slices by a factor of 10 or more.
From Eq.~(\ref{final}) we can note, that the first free-particle terms can be considered as a weight over all possible random walks (Brownian random walks) in the imaginary time $\beta$ with the ends points $R$ and $R'$. In the limit $M\rightarrow \infty$ we directly obtain the Feynman-Kac relation
\begin{equation}
\rho(R,R'; \beta)=\rho_0(R,R'; \beta) \left\langle e^{-\int\limits_0^{\beta} dt \,V(R(t))}\right\rangle_{FK}.
\label{FK}
\end{equation}
In the quasi-classical limit ($\beta \rightarrow 0$), only the classical path is important,
$
R_0(t)=\left(1 -\frac{t}{\beta} \right)R+ \frac{t}{\beta} R',
$
which leads to the semi-classical approximation of the high-temperature density matrix
\begin{equation}
\rho(R,R';\Delta \beta)=\rho_0(R,R'; \Delta \beta) e^{-\int\limits_0^{\Delta\beta} dt \,V(R_0(t))},
\end{equation}
which is already much better compared to Eq.~(\ref{coord_pr}) with the substitution of classical (in many cases divergent) potentials.

For systems with pair interactions, in the limit of small $\Delta \beta$, the full density matrix~(\ref{FK}) can be approximated by a product of pair density matrices
\begin{equation}
\left\langle e^{-\int\limits_0^{\beta} dt \,V(R(t))}\right\rangle_{FK} \approx \prod\limits_{j < k} \left\langle e^{-\int\limits_0^{\beta} dt \,V_{pair}[\mathbf r_j(t),\mathbf r_k(t)]}\right\rangle_{FK}.
\label{BW}
\end{equation}

This is used in practice as the {\rm pair approximation}. It supposes that on the small time interval $\Delta \beta$ the correlations of two particles become independent from the surroundings. Different derivations of the effective pair potential (average on the l.h.s of Eq.~(\ref{BW})) have been proposed in literature~\cite{kelbg,fil_pre,Kleinert2}.

Implementation of the periodic boundary conditions leads to further modifications, see e.g. Refs.~\cite{Gaskell,cep01,kent99}.

{\em \textbf{C. How to calculate physical properties}}\\
There are different approaches for calculating expectation values of physical observables, such as the energy, momentum distribution, etc., which are called estimators. In each particular case convergence can be improved by choosing the proper estimator.
Consider, for example, the thermodynamical estimator of the internal energy
\begin{equation}
E=-\frac{\partial}{\partial \beta} \left(ln Z\right)=- \frac{1}{Z}
\frac{\partial Z}{\partial \beta}=-\frac{1}{Z}\int dR \frac{\partial \rho(R,R;\beta)}{\partial \beta}.
\end{equation}
After direct substitution of Eq.~(\ref{final}) for $\hat \rho$, one obtains\footnote{This  is only valid for particles with Boltzmann statistics. For fermions one has to include additional terms related to the $\beta$-derivative of the exchange determinant~\cite{fil_contrib,numbook06}.}
\begin{equation}
E= \frac{d M N}{2\beta} - \left\langle \sum_{i=0}^{M-1}
\frac{M m}{2 \hbar^2 \beta^2} (R_{i}-R_{i+1})^2
\right\rangle_{\rho(R;\beta)}+\left\langle \frac{1}{M}\sum_{i=0}^{M-1} V(R_i)
\right\rangle_{\rho(R;\beta)}.
\label{direct}
\end{equation}
This form of the energy estimator has a much larger statistical variance $\sigma_s$ compared to the virial estimator~\cite{herman}. Since the statistical error in Monte Carlo simulations decreases as $\delta E\approx\sigma_s/\sqrt{N_{MC}}$ (with $N_{MC}$ being the number of MC-steps), with the direct estimator~(\ref{direct}) one usually needs several times more MC runs to get the same accuracy as given by the virial estimator.

One of the approaches to obtain the virial estimator of energy  is to introduce temperature dependent coordinates~\cite{fil_fusion}, i.e.
$
\tilde R_{i}= R_{0}+\lambda_{\Delta} \sum_{m=1}^{i} \xi^{m}, \;
i=1,\ldots,M-1. \label{trans}
$
Here $\xi^{i}$ is a set of unit vectors, and $R_{0}$ is a set of particle coordinates at the zero time slice ($R_0=R$). Once this has been done, the estimator takes the form~\cite{numbook06}
\begin{equation}
E=\frac{d N}{2\beta}+ \left\langle V(\tilde R) +  \beta\frac{\partial V(\tilde
R)}{\partial \tilde R} \frac{\partial \tilde R}{\partial
\beta} \right\rangle_{\rho(\tilde R;\beta)}.
\label{virial}
\end{equation}
One can note at once, that for weakly interacting systems at high temperatures, the virial result~(\ref{virial}) directly gives the classical kinetic energy (first term) and does not depend on the chosen number of time slices $M$, whereas in the direct estimator~(\ref{direct}) we get this result by calculating the difference of two large terms which are diverging as $M \rightarrow \infty$.

{\em \textbf{D. Acceptance Ratio}}\\
When we try different kinds of moves in the Metropolis algorithm, it can turn out that some moves will be frequently rejected or accepted. In both cases, we loose the efficiency of the algorithm. For a sufficiently long time (number of MC steps) we will be trapped in some local region of phase space and will not explore the whole space in a reasonable computer time. In practice, the parameters of the moves are usually chosen to get an acceptance ratio of roughly $50 \%$. For different kinds of moves in PIMC (particle displacement, path deformation, permutation sampling) such an acceptance ratio is preferable and construction of good apriori sampling distributions is crucial.

A detailed discussion of these topics is beyond the scope of the present paper and is covered in other publications, e.g. Refs.~\cite{Cep95,numbook06}.

{\em \textbf{E. Quantum exchange. PIMC for bosons/fermions}}\\
Now we come to ``real'' quantum particles. As we have already discussed, properties of a system of $N$ particles at a finite temperature $T$ are determined by the density operator. Due to the Fermi/Bose statistics the total density matrix should be (anti)symmetric under arbitrary exchanges of identical particles (e.g. electrons, holes, with the same spin projection), i.e. we have to replace  ${\hat{\rho }\rightarrow {\hat{\rho }}^\mathrm{A/S}}$ for fermions/bosons. As a result the full density matrix will be a superposition of all $N!$ permutations of $N$ identical particles. Consider the case of two types (e,h) of particles with numbers $N_e$, $N_h$
\begin{equation}
\rho^\mathrm{A/S}(R_\mathrm{e},R_\mathrm{h},R_\mathrm{e},R_\mathrm{h};\beta)=\frac{1}{N_{e}!N_{h}!}\sum _{P_\mathrm{e}P_\mathrm{h}}(\mp 1)^{P_\mathrm{e}}(\mp 1)^{P_\mathrm{h}}\rho(R_\mathrm{e},R_\mathrm{h},\hat{P}_\mathrm{e}R_\mathrm{e},\hat{P}_\mathrm{h}R_\mathrm{h};\beta),
\label{rho_as}
\end{equation}
where $P_{e(h)}$ is the parity of a permutation (number of equivalent pair transpositions) and $\hat{P}_\mathrm{e(h)}$ the permutation operator. We directly see that, for bosons all terms have a positive sign, while for fermions the sign of the prefactor alternates depending whether the permutation is  even or odd.

In the last case a severe problem arises. The Metropolis algorithm gives the same distribution of permutations for both Fermi and Bose systems. The reason is that, for sampling permutations, we use the modulus of the off-diagonal density matrix, $\vert \rho(R,\hat P R; \beta) \vert$.
\begin{itemize}
\item Bose systems: all permutations contribute with the same (positive) sign. Hence with the increase of the permutation statistics, accuracy in the calculation of the density matrix increases proportionally.
\item Fermions: essential cancellation of positive and negative terms (corresponding to even and odd permutations), both are close in their absolute value. Accurate calculation of this small difference is drastically hampered with the increase of quantum degeneracy (low T, high density). The consequences are large fluctuations in the computed averages. This is known as the {\em fermion sign problem}. It was shown~\cite{fixed3} that the efficiency of the straightforward calculations scales like $e^{-2 N \beta \Delta F}$, where $\Delta F$ is the free energy difference per particle of the same fermionic and bosonic system, and $N$ is the particle number.
\end{itemize}

%As an illustration let us consider $2$ electrons and $2$ holes (fermions)
%\[\rho ^{A}(R_{e},R_{h},R_{e},R_{h};\beta) = \frac{1}{2!2!}\sum_{P_{e}P_{h}}(-1)^{P_{e}}(-1)^{P_{h}}\rho(R_{e},R_{h},\hat{P}_{e}R_{e},\hat{P}_{h}R_{h};\beta)\]

{\em \textbf{F. Numerical sampling of permutations}}\\
Fermi and Bose statistics require sampling of permutations, see Eq.~(\ref{rho_as}), in addition to the integrations in real space. From the $N!$ possibilities, we need to pick up a permutation which has a non-zero probability for a given particle configuration.

To realize a permutation we pick up, along the $\beta$-axis, two end-points $\{R_{i}, R_{i+i_0}\}$ with $i_0=2^{l-1}\,  (l=1,2,\ldots)$. Although the permutation operator $\hat P$ in Eq.~(\ref{rho_as}) acts on the last time-slice, $R_{e(h)}$, the permutation of the paths, $\{R_{i}, R_{i+i_0}\} \rightarrow \{R_{i}, \hat P R_{i+i_0}\}$ can be carried out at any time slice because the operator $\hat P$ commutes with the Hamiltonian. In a permutation move ($k$ permuted particles) the path coordinates between the fixed points $R_{i}$ and $R_{i+i_0}$ are removed and new paths connecting one particle to another (new $k$ links) or a new path connecting a particle on itself (if a given particle undergoes the identity permutation) are sampled.

It is evident, that a local permutation move consisting of a cyclic exchange of $k \geq 2$ neighboring particles will be more probable than a global exchange involving a macroscopic number of particles, and, in general, the probability of exchange will decrease with the increase of $k$. The most probable are local updates: permutations of only few ($2,3,4$) particles. Moreover, any of the $N!$ permutations can be decomposed in a sequence of successive pair transpositions (two particle exchange), and we can explore the whole permutation space by making only local updates which have a high acceptance ratio.

In MC simulations we choose as the sampling probability of permutations
\begin{equation}
T(P\rightarrow P')=\frac{\rho_{kin}(R_i,\hat P R_{i+i_0};i_0 \Delta \beta)}{\sum\limits_{P\in \Omega(P)} \rho_{kin}(R_i,\hat P R_{i+i_0};i_0 \Delta \beta)},
\end{equation}
where we have used the product of the $k$ one-particle density matrices
\begin{equation}
\rho_{kin}(R_i,\hat P R_{i+i_0},;i_0 \Delta \beta)\propto \exp\left( -\sum_{j\in k} \pi (\mathbf r_{i}^j- \hat P \mathbf r_{i+i_0}^j)^2/(i_0 \lambda^2_{\Delta})\right).
\end{equation}
Here $\Omega (P)$ denotes the neighborhood of the current permutation $P$ from which the permutation $P'$ is sampled. For example, for the two-particle exchange, $\Omega (P)$ equals to number of neighbors of the given particle in the range of several De Broglie wavelengths, $\lambda(t),\, t=i_0 \Delta \beta$, which are possible candidates for the exchange.

To satisfy the detailed balance principle, we make a final decision\footnote{The sampled permutation can be rejected earlier when the new paths connecting $R_i$ and $\hat P R_{i+i_0}$ are sampled with the bisection algorithm~\cite{Cep95,numbook06}.} about the sampled permutation using the acceptance probability
\begin{equation}
A(P\rightarrow P')=\mbox{min}\left[1, \frac{\sum\limits_{P\in \Omega(P)} \rho_{kin}(R_i,\hat P R_{i+i_0};i_0 \Delta \beta)}{\sum\limits_{P'\in \Omega(P')} \rho_{kin}(R_i,\hat P' R_{i+i_0};i_0 \Delta \beta)} \right],
\end{equation}
where $\Omega(P')$ is the neighborhood of the new permutation $P'$. If the neighborhoods of the current and new permutation are equal, the acceptance probability is one.

As an illustration, in Fig.~\ref{bisec_exam} we show a {\em world line picture} of five particles. Particle indices in Fig.~\ref{bisec_exam}~(a) and~(c) are placed near the starting and end point of the particle trajectories. Hence, when the sequence of indices at $m=0$ and $m=100$ does not coincide, see Figs.~\ref{bisec_exam}~(a)(b), the particles are permuted.

As we can see from Fig.~\ref{bisec_exam}(a) the paths of particles ``$1$'' and ``$2$'' are closed (two identical permutations), three other particles are in one cyclic exchange, and the whole permutation  can be denoted as $\{1,2,5,3,4\}$ (as we can see the end of the path ``$3$'' coincides with the beginning of path ``$5$'', the end of path ``$4$'' coincides with the beginning of path ``$3$'' and the path ``$5$'' ends up at the starting position of path ``$4$''. Now we decide to make a transposition between particles ``$1$'' and ``$4$''. To do this we choose randomly time slices where new paths will be sampled. In our case it was $m=17-33$. First, we exchange the edge points at the time slice $m=33$, i.e. $\mathbf r_{1}'^{33}\equiv \mathbf r_{4}^{33}$ and $\mathbf r_{4}'^{33}\equiv \mathbf r_{1}^{33}$. Hence the position of the edge points is not sampled, they are a part of the unchanged trajectories. Once the initial and final points are chosen, we use the bisection algorithm to sample two paths connecting edge points, see Fig.~\ref{bisec_exam}(b).

\section{PIMC for degenerate Bose systems}
\label{bosons}

Currently much experimental activity is devoted to the study of ensembles of dilute gases of Bose atoms and optically excited indirect excitons in single/double well nanostructures, e.g. Refs.~\cite{negoita99,kim04} and references therein. The most exciting is certainly the possibility to observe signatures of Bose condensation and superfluidity. The essential point of these experiments is that the number of trapped atoms are limited to a few ten thousand particles and one should expect significant deviation (finite size effects) from the macroscopic limit, leading e.g. to a ``softening'' of the condensate fraction curve in the transition region and also to a shift of the critical temperature to lower values. This is particularly important for the case of a few hundreds of particles which become accessible to a direct theoretical investigation using Quantum Monte Carlo approaches which allows to treat many-body correlation effects from ``first principles''.
% existing quantum mechanical
%simulation techniques.

In PIMC, as was shown by Feynman~\cite{Feynm}, the Bose statistics manifest itself as a special topology of the particle trajectories which can form macroscopically large permutation cycles. The free external parameters, like temperature, density, interaction strength, have a direct influence on these cycle distributions and, hence, on the superfluid and condensate fractions. Below we demonstrate how the latter can be easily related to the statistics of path configurations sampled by PIMC.

To be more specific in the discussion below we consider a system of trapped bosons with Coulomb interaction described by the Hamiltonian (which can be also reduced to the dimensionless form, $\tilde H$, using the system of units: $r \rightarrow \tilde r=(r/a_B), E \rightarrow \tilde E=(E/\text{Ha})$ with $a_B=\hbar^2\epsilon/m e^2$ and $\text{Ha}=e^2/\epsilon a_B$)
\begin{eqnarray}
\hat H &=& \hat H_{o} + \sum\limits_{i< j }^{N} \frac{e^2 }{\varepsilon \vert {\bf r}_i- {\bf r}_j\vert^2 } , \quad
\hat H_o = \sum\limits_{i=1}^{N} \left( -\frac{\hbar^2}{2 m} \nabla_{{\bf r}_i}^2 + \frac{m}{2}\omega^2 r_i^2 \right), \label{Ham} \\
\tilde H &=&\hat H/\text{Ha}= \sum\limits_{i=1}^{N} \frac 1 2\left(- \nabla_{\tilde{\bf r}_i}^2 + \frac{1}{\lambda^{4}} \tilde r_i^2 \right) + \sum\limits_{i< j }^{N} \frac{1}{\tilde r_{ij}}.
\label{Ham2}
\end{eqnarray}
For the mesoscopic trapped system as considered here the density is controlled by the
harmonic trap frequency $\omega$ and is characterized by the coupling parameter
$\lambda=(e^2/\epsilon l_0)/(\hbar \omega)=l_0/a_B$ with
$l_0^2=\hbar/m \omega$. In the limit $(\lambda \rightarrow \infty)$ the external potential
vanishes, while for $(\lambda \rightarrow 0)$ the Coulomb interaction can be neglected (formal transition to non-interacting bosons).

{\em \textbf{A. Superfluidity}}\\
First, we consider the fraction of the superfluid (mass) density $\gamma_s=\rho_s/\rho$ which, within the Landau two-fluid model is computed from the classical and quantum momenta of inertia, $I_c$ and  $I_q$, according to $\gamma_s=1-I_q/I_c$~\cite{landau}. The quantities $I_c$ and  $I_q$ are effectively computed in PIMC simulations~\cite{Sing} from the area enclosed by the particle paths $\bf A$, using
\begin{equation}
\frac{\rho_s}{\rho}= \frac{4 m^2 \langle A_z^2 \rangle}{\hbar^2 \beta \langle I_z \rangle},\;\; \mathbf A= \frac 1 2 \sum_{i=1}^{N} \sum_{k=0}^{M-1} \mathbf r^i_{k} \times \mathbf r^i_{k+1}, \quad
I_c=\left\langle  \sum_{i=1}^{N} \sum_{k=0}^{M-1} m_i  \mathbf r^{i}_{k, \perp} \cdot \mathbf r^{i}_{k+1, \perp} \right\rangle
\label{iq}
\end{equation}
where $N$ is the particle number, $M$ the number of time slices used in the path integral presentation and $\langle \ldots\rangle$ denotes the thermal averaging with the symmetric $N-$particle diagonal density matrix, i.e.
\begin{equation}
\langle \ldots \rangle =\frac 1 Z \int \int d\mathbf r_1 d\mathbf r_2 \ldots d\mathbf r_N
\; (\ldots)\; \rho^S(\mathbf r_1,\mathbf r_2,\ldots,\mathbf r_N; \beta).
\end{equation}
This formula has been derived~\cite{Sing} for {\em finite systems} by assuming that particles are placed in an external field, e.g. a rotating cylinder. Then one considers that the system is put in a permanent slow rotation with the result that the normal component follows the imposed rotation while the superfluid part stays at rest. The effective moment of inertia is defined as the work done to rotate the system by a unit angle.

For macroscopic systems the path area formula~(\ref{iq}) can be modified~\cite{Cep95,Poll}. Instead of a filled cylinder, one considers two cylinders with the radius $R$ and spacing $d$, where $d \ll R$. Such a torus is topologically equivalent to the usual periodic boundary conditions. As a result we have: $I_c=m N R^2$ and $A_z=WR/2$, where $W$ is the {\em winding number}, defined as the flux of paths winding around the torus and crossing any plane
\begin{equation}
\gamma_s=\frac{\rho_s}{\rho}= \frac{\langle W^2 \rangle}{2 \lambda \beta N}, \quad \mathbf W= \sum_{i=1}^N \int_0^{\beta} dt \left[ \frac{d \mathbf r_i(t)}{dt} \right].
\end{equation}

{\em \textbf{B. Off-diagonal long-range order}}\\
The magnitude of off-diagonal long-range order is, in macroscopic systems, also directly accessible with PIMC. It is characterized by the asymptotic behavior of the single-particle off-diagonal density matrix
\begin{align}
\rho(\mathbf r_1, \mathbf r'_1;\beta) &=
	V/Z \int d \mathbf r_2 \ldots  d \mathbf r_N \; \rho^S(\mathbf r_1,\mathbf r_2,\ldots,\mathbf r_N,\mathbf r'_1,\mathbf r_2,\ldots,\mathbf r_N; \beta),\label{rho_singl} \\
n_0(\beta) &= \lim\limits_{\mathbf r'_1\rightarrow \infty} \rho(\mathbf r_1, \mathbf r'_1;\beta),
\label{rho_off}
\end{align}
where $n_0$ is the fraction of particles in the condensate and $V$ is the volume of the simulation cell. For a homogeneous isotropic system, $\rho(\mathbf r_1, \mathbf r'_1)=\rho(|\mathbf r_1- \mathbf r'_1|)$ and, by taking the Fourier transform of an off-diagonal element, one obtains the momentum distribution
\begin{equation}
\rho(\mathbf k)=\frac{1}{(2\pi)^{d}} \int d(\mathbf r_1 -\mathbf r'_1) \, e^{-i \mathbf k (\mathbf r_1-\mathbf r'_1)}\; \rho(|\mathbf r_1- \mathbf r'_1|; \beta),
\label{p_dis}
\end{equation}
which shows a sharp increase at zero momentum when the temperature drops below the critical temperature $T_c$ of Bose condensation.

Obviously, a finite trapped system of particles considered in real experiments behaves differently. The radial density is strongly inhomogeneous with the highest value at the trap center. However, these systems do represent an analog of the homogeneous macroscopic system in the {\em angular} direction (for traps with angular symmetry as in the case~(\ref{Ham})). Hence, the macroscopic formulas~(\ref{rho_off}) and~(\ref{p_dis}) should be modified in an appropriate way and the corresponding momentum distribution, the condensate fraction and superfluidity get an additional dependence on the radial distance from the trap center.

As follows from Eq.~(\ref{rho_singl}), for the numerical evaluation of the single-particle density matrix one should allow that one of the $N$ simulated particles has an open path, e.g. $\mathbf r_1 \neq \mathbf r'_1$. The paths of the other $N-1$ particles can close on their beginning (identical permutation) or at the start of another particle's path. The coordinates $\mathbf r_1, \mathbf r'_1$ are independent but their probability is given by the $N$-particle density matrix. In simulations we record a histogram (distribution) given by
\begin{gather}
\rho(\mathbf r, \mathbf r';\beta) \propto \left\langle \delta(\mathbf r_1 -\mathbf r) \delta (\mathbf r'_1 -\mathbf r')\right\rangle_{W},\\
W = \rho^S(\mathbf r_1,\mathbf r_2,\ldots,\mathbf r_N,\mathbf r'_1,\mathbf r_2,\ldots,\mathbf r_N; \beta)/Z',
\end{gather}
($Z'$ is the normalization factor) which is then used to obtain the momentum distribution~(\ref{p_dis}). The probability $W$ is sampled using the path integral representation of $\rho^S$.

Recently a new method to sample the single-particle density matrix~(\ref{rho_singl}), by generalization of the conventional PIMC to the grand canonical ensemble, has been proposed~\cite{wa_first}. A {\em worm algorithm} allows for a simultaneous sampling of both diagonal configurations contributing to the partition function and off-diagonal ones which contribute to the one-particle Matsubara Green function. The method has been recently applied to study of Bose condensation in crystalline $^4He$ and  superfluidity in para-hydrogen droplets~\cite{wa_app}, where high efficiency in sampling of long permutation cycles (practically unaffected by system size) and significantly improved  convergence in the calculation of superfluid properties has been demonstrated.

\section{Numerical results}
{\em \textbf{A. Reference case: Ideal system}}\\
For the noninteracting Bose gas the total Hamiltonian is a sum over independent single particle Hamiltonians, $\hat H =\hat H_o$, Eq.~(\ref{Ham}). If we then calculate the partition function as the trace over the symmetrized density matrix $\rho^S(R,R;\beta)$, it can be written as a product of $N$ independent integrals
\begin{equation}
Z_N(\beta)=\frac{1}{N!}\sum_{P\in S_N}\prod_{i=1}^N \int \mathrm{d}{\mathbf r^i}\,\rho_1(\mathbf r^i,\hat P\mathbf r^{i};\beta),
\end{equation}
where each integral contains only the single-particle density matrix.

Any permutation can be broken into exchange cycles. A given permutation consists of a set of cycles $\{C_1,C_2,\ldots,C_N\}$, where $C_q$ denotes the number of cycles with length $q$. As the particles are unlabeled, different permutations resulting in identical cycle configurations give identical contributions to the partition function. We can thus rewrite the sum over permutations as a sum over cycle configurations with an additional factor taking into account the number of possible realizations. By applying the group property of the density matrix the integrals over the single-particle density matrices within a cycle of length $q$ are reduced to a single integral over the density matrix at the $q$-times lower temperature with the result~\cite{FeynmStat}
\begin{equation}\label{f:5.ZCq}
Z_N(\beta)=\sum_{\substack{\{C_q\}\\ \text{restr.}}} \prod_{q=1}^N \frac{Z_1(q\beta)^{C_q}}{C_q!q^{C_q}},
\end{equation}
with the restriction $\sum_{q=1}^{N} qC_q=N$, for the possible cycle configurations.

The same procedure can be applied for the calculation of any thermal average. In particular, for the mean cycle occupation number $\avr{C_q}$ we obtain
\begin{equation}\label{f:5.Cqrec}
\avr{C_q} =\frac{1}{Z_N(\beta)}\sum_{\substack{\{C_r\}\\ \text{restr.}}}\prod_{r=1}^N \frac{Z_1(r\beta)^{C_r}}{C_r!r^{C_r}}C_q=\frac{Z_{N-q}(\beta)}{Z_N(\beta)}\frac{Z_1(q\beta)}{q}.
\end{equation}
Plugging this formula into the constraint on the possible cycle configurations leads to a powerful recursion relation for the partition function~\cite{Kleinert,Weiss97}
\begin{equation}\label{f:5.Zrec}
Z_N(\beta) =\frac{1}{N}\sum_{q=1}^N Z_{N-q}(\beta) Z_1(q\beta).
\end{equation}
For the energy of the system we obtain ($\tau=1/\beta$)
\begin{equation}\label{f:5.Eideal}
 \avr{E}=\tau^2\frac{\partial\ln Z_N(\tau)}{\partial\tau}=\sum_{q=1}^N E_1(q\beta)\avr{qC_q}.
\end{equation}

The occupation number $N_i$ of an arbitrary energy level $E_i$ can be calculated as the derivative of the partition function with respect to $\beta E_i$, see Ref.~\cite{Weiss97}
\begin{equation}
\avr{N_i} = -\frac{1}{Z_N(\beta)}\frac{\partial Z_N(\beta)}{\partial(\beta E_i)}=\sum_{q=1}^{N}\frac{e^{-q\beta E_i}}{Z_1(q\beta)}\avr{qC_q}.
\end{equation}
This formula yields the number of particles in the condensate, when taken at the ground state energy $E_0$.

Within linear response theory, the quantum mechanical moment of inertia $I_\mathrm{q}$ is defined as the system response to rotations. A careful analysis for non-rotating situations with $\avr{L_z}=0$ of a system confined in a 2d spherical harmonic trap yields~\cite{Schneider00}
\begin{equation}
I_\mathrm{q}=\beta\avr{L_z^2}=2\hbar^2\sum_{q=1}^N \frac{q\beta e^{-q\beta\hbar\omega}}{(1-e^{-q\beta\hbar\omega})^2} \avr{qC_q}.
\label{Iq}
\end{equation}
The classical moment of inertia $I_\mathrm{c}$ of the same system becomes
\begin{equation}
I_\mathrm{c}=m\sum_{j=1}^{N}\avr{x_j^2+y_j^2}=\frac{\hbar}{\omega}\sum_{q=1}^N\frac{1+e^{-q\beta\hbar\omega}}{1-e^{-q\beta\hbar\omega}} \avr{qC_q}.
\label{Ic}
\end{equation}

\begin{figure}[t]
\vspace{-0.8cm}
\centering\includegraphics[width=0.7\textwidth,angle=-90]{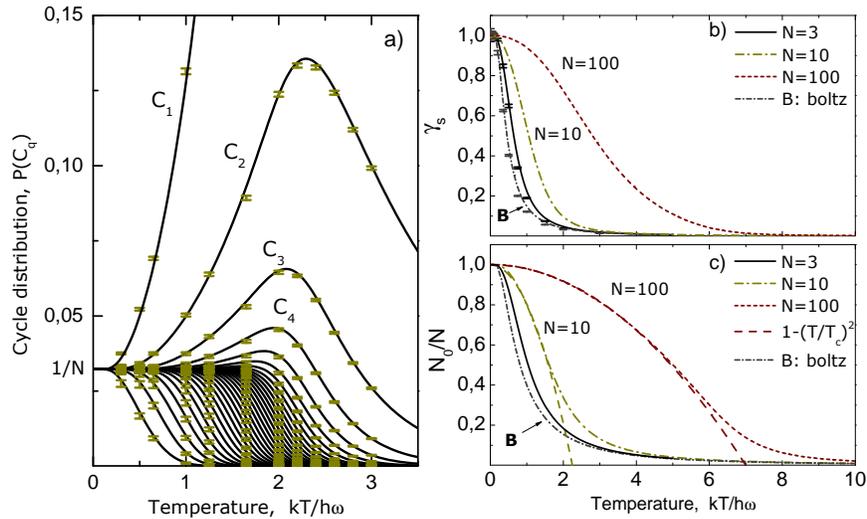}
\vspace{-0.60cm}
\caption{a) Cycle distribution for 31 non-interacting bosons in a 3D harmonic trap; 
Symbols with error bars denote PIMC results. Analytical results, Eq.~(\ref{f:5.Cqrec}), are shown with lines. $C_q$ denotes the permutations of length $q$.
Right panel: Superfluid fraction (b) and condensate fraction (c) for varying  numbers of 2D trapped bosons. Results for boltzmannons (``boltz'', B) do not depend on particle number $N$.}
\label{fig:ideal_cyc_superf}
\end{figure}

The left side of Fig.~\ref{fig:ideal_cyc_superf} shows a typical cycle configuration, i.e. the probability, that a particle belongs to an exchange cycle of length $q$. Clearly, analytical and numerical results are in good agreement. Both indicate that every possible permutation is equally probable at low temperatures, whereas only cycles of length~$1$ occur at higher temperatures (uppermost curve $C_1$). In this regime Bose statistics are of no importance and a PIMC simulation without permutation sampling results in the same values for thermal quantities. Note, that this does not necessarily imply that the system can be treated classically. It rather means that the single-particle wave-functions have a reasonably small overlap, while still extending over a finite space. Such quantum particles are called ``Boltzmannons''.

The effect of Bose statistics on thermal averages at low temperatures can be further investigated by switching it on and off in the PIMC program. In the analytical formulas for the noninteracting Bose gas this is easily achieved by setting $C_1=N$ and putting every longer cycle number to zero. The impact on the superfluid fraction $\gamma_s$ is shown in Fig.~\ref{fig:ideal_cyc_superf}(b). In both cases it reaches unity at absolute zero and disappears for high temperatures. The fact that a noninteracting system shows a superfluid response may seem surprising since superfluidity is a ground state property which appears in systems with a linear dispersion and is, thus, inseparably connected with inter-particle interactions. However, it does occur even in mesoscopic ideal systems due to the discrete nature of the excitation spectrum in finite systems, leading to energy gaps.

Fig.~\ref{fig:ideal_cyc_superf}(b) indicates that in a finite trapped system
superfluidity appears also for boltzmannons.
Nevertheless, particle exchange allows condensation into the ground state at higher temperatures which ultimately increases the superfluid fraction for all temperatures. In contrast to the boltzmannonic calculations, the superfluid fraction in the bosonic system, therefore, depends on the particle number. An explanation for this behavior can be drawn from the analytical expressions for the thermal averages. These depend only on single-particle quantities as it should be the case for an ideal system. But an exchange cycle of length $q$ contributes to an average like a single particle at a $q$-times lower temperature. The path-integral method uses the same principle in the opposite direction---a single-particle is described as a cycle.

\begin{figure}[t]
\vspace{-1.0cm}
\includegraphics[width=0.75\textwidth,angle=-90]{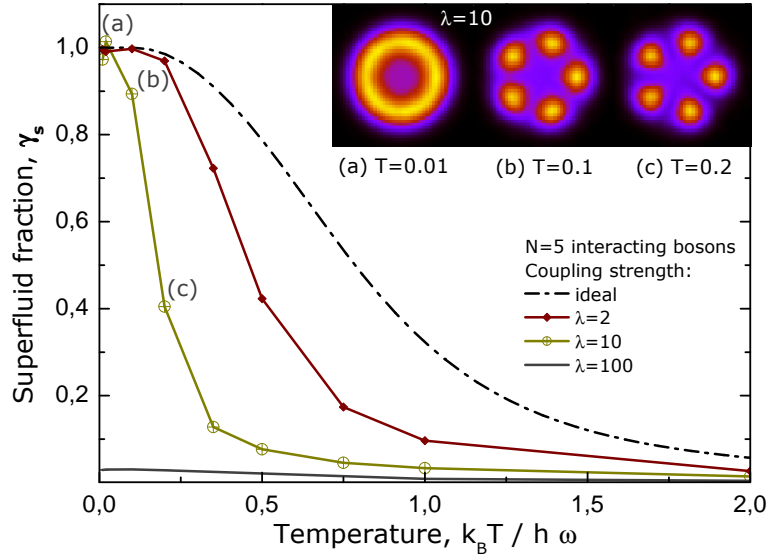}
\vspace{-0.6cm}
\caption{Superfluid fraction for 5 interacting bosons, $0 \leq \lambda \leq 100$, in a 2D harmonic trap versus temperature, $T=k_B T/\hbar \omega$. Symbols denote PIMC simulations. Dashed-doted  line displays an analytical result, $\gamma_s=1-I_q/I_c$, for ideal system, Eqs.~(\ref{Iq}-\ref{Ic}). The inserts show the density distributions at $\lambda=10$ and three temperatures: denoted as (a),(b),(c).}
\label{fig:inter_superf_N5}
\end{figure}

Condensate and superfluid fraction are not identical as it can be seen in Fig.~\ref{fig:ideal_cyc_superf}(b),(c). The higher the particle number, the more the condensate fraction shows its typical polynomial dependence on the temperature (which is $1-(T/T_\mathrm{c})^2$ for 2D trapped systems where $T_c$ is the critical temperature). In the thermodynamic limit the ``softening'' at the transition point vanishes completely. The system then has a clearly defined critical temperature below which Bose-Einstein condensation occurs. On the other hand, taking the thermodynamic limit means an infinite system volume which implies $\hbar\omega\to0$. Thus, the superfluid density will disappear completely. The plots in 
Fig.~\ref{fig:ideal_cyc_superf} cannot show this behavior due to the chosen temperature scale which keeps $\hbar\omega$ constant.

{\em \textbf{B. Interacting Bosons}}\\
With the PIMC method it is possible to include inter-particle interactions like e.g. Coulomb repulsion~(\ref{Ham}) from first principles. The effective strength of the interaction can be controlled by the trap frequency and is measured by the parameter $\lambda$. Fig.~\ref{fig:inter_superf_N5} shows the temperature dependence of the superfluid fraction for several values of $\lambda$ (e.g. $2 \leq \lambda \leq 10$ corresponds to typical particle densities in semiconductor quantum dots). The repulsive interaction causes a shift of the transition temperatures to lower values. When cooled down, the system typically forms a crystal like state in intermediate temperature regions until it melts into a ring like structure with delocalized particles, see insets in Fig.~\ref{fig:inter_superf_N5}. Obviously, the latter shows a high superfluid response which is proportional to the ratio of the area enclosed by paths to the cross-section of the whole system (see Eq.~\ref{iq}). In the ideal case, the system skips the intermediate crystal phase and directly reaches the delocalized state. In the case of dominating interaction strengths, the system stays highly localized even at absolute zero. Note, that even for the crystal phase the simulations yield a non vanishing value $\gamma_s$. This is a finite size effect because of a nonzero area ratio~(\ref{iq}).

\section{Discussion}

a) Let us summarize the main areas of application of the PIMC method:
\begin{itemize}
\setlength{\itemsep}{0.1mm plus3mm minus1mm}
\item Low temperature systems (relevance of quantum effects).
\item Small dimensions (system size comparable to De Broglie wavelength $\lambda_D$).
\item High density: 2-particle distance comparable to or / smaller than $\lambda_D$.
\end{itemize}

\noindent b) The simulations allow to solve problem of $1,2$ or $N$ quantum particles:
\begin{itemize}
\setlength{\itemsep}{0.1mm plus3mm minus1mm}
\item Single particle in a complicated potential, e.g. disorder effects $\rightarrow$ effective solution of Schr\"odinger's equation  (ground state, $T=0$) or finite temperature extension (density matrix)~\cite{numbook06}.
\item Improvement of pair potential at small distances (include quantum effects)~\cite{fil_pre}.
\item Finite number of particles in traps atoms, ions at ultra low temperature (Bose condensates etc.) electrons, holes in quantum dots, e.g.~\cite{prl01}.
\item Macroscopic quantum systems of electrons and ions in astrophysics: planet cores, dwarf stars, highly excited semiconductors (many electrons, holes in nanostructures), e.g.~\cite{fil_fusion,fil_contrib}.
\end{itemize}

\noindent c) Naturally, this list is far from being complete. It is the generality of this first principle approach which makes it increasingly important for many fields of physics and beyond.


\begin{thebibliography}{10}
\setlength{\itemsep}{0.1mm plus3mm minus1mm}    % Squeeze bibitem separation a bit

\bibitem{other} See the chapters on classical Monte Carlo.

\bibitem{Feynm} R.P.~Feynman and A.R.~Hibbs, {\em Quantum Mechanics and Path
Integrals}, McGraw Hill, New York 1965.

\bibitem{FeynmStat} R.P.~Feynman {\em Statistical Mechanics}, W. A. Benjamin, Inc., Advanced Book Program, Reading, Massachusetts 1972.

\bibitem{Kleinert}H.~Kleinert, {\em Path Integrals in Quantum Mechanics, Statistics
and Polymer Physics}, World Scientific, Second edition, 1995.

\bibitem{fixed3}
D.M.~Ceperley, {\em Path Integral Monte Carlo Methods for Fermions}
in {\em Monte Carlo and Molecular Dynamics of Condensed Matter Systems},
Ed. K. Binder and G. Ciccotti, Editrice Compositori, Bologna, Italy, 1996.

\bibitem{Cep95} D.M.~Ceperley, Rev. Mod. Phys. {\bf 67}, 279 (1995).

\bibitem{numbook06} A.~Filinov and M~Bonitz, in: {\em Introduction to Computational Methods for Many Body Systems}, M.~Bonitz and D.~Semkat (eds.), Rinton Press, Princeton 2006.

\bibitem{kelbg} W.~Ebeling, H.J.~Hoffmann, and G.~Kelbg, Contr. Plasma Phys. \textbf{7},
233 (1967) and references therein.

\bibitem{fil_pre} A.V.~Filinov, V.O.~Golubnychiy, M.~Bonitz,
 W.~Ebeling, and J.W.~Dufty, Phys. Rev. E {\bf 70}, 046411 (2004).

%\bibitem{storer} R.G.~Storer, J. Math. Phys. \textbf{9}, 964 (1968); A.D.~Klemm,
%and R.G.~Storer, Aust. J. Phys. \textbf{26}, 43 (1973).

\bibitem{Kleinert2} H.~Kleinert, Phys. Rev. D \textbf{57}, 2264
(1998).

\bibitem{Gaskell} T.~Gaskell, Proc. Phys. Soc. {\bf 77} 1182 (1961); {\bf 80}, 1091 (1962); D.M.~Ceperley, Phys. Rev. B {\bf 18}, 3126 (1978); V.~Natoli and D.M.~Ceperley, J. Comput. Phys. {\bf 117}, 171
(1995).

\bibitem{cep01} C.~Lin, F.H.~Zong, and D.M.~Ceperley, Phys. Rev. E {\bf 64}, 016702
(2001).

\bibitem{kent99} P.R.C.~Kent, R.Q.~Hood, A.J.~Williamson, R.J.~Needs, W.M.C.~Foulkes,
and G.~Rajagopal, Phys. Rev. B {\bf 59}, 1917 (1999);
A.J.~Williamson, G.~Rajagopal, R.J.~Needs, L.M.~Fraser, W.M.C.~Foulkes,
Y.~Wang, and M.-Y.~Chou, Phys. Rev. B {\bf 55}, R4851.

\bibitem{herman}
M.F.~Herman, E.J.~Bruskin, and B.J.~Berne,
J. Chem. Phys. {\bf 76}, 5150 (1982).

\bibitem{fil_fusion}
V.S.~Filinov, M.~Bonitz, W.~Ebeling, and V.E.~Fortov,
Plasma Physics and Controlled Fusion {\bf 43}, 743 (2001).

\bibitem{fil_contrib} V.S.~Filinov, M.~Bonitz, D.~Kremp, W.D.~Kraeft, W.~Ebeling, P.R.~Levashov, and V.E.~Fortov, Contrib. Plasma Phys. {\bf 41}, 135 (2001).

\bibitem{negoita99} V. Negoita, D.W. Snoke, and K. Eberl, Phys. Rev. B {\bf 60}, 2661 (1999); L.V. Butov et al., Phys. Rev. Lett. {\bf 86}, 5608 (2001).
\bibitem{kim04} E. Kim, and M. Chan, Nature (London) {\bf 427}, 225 (2004)

\bibitem{landau} L.D.~Landau, J. Phys. USSR {\bf 5}, 71 (1941); E.L.~Andronikashvili, J. Phys. USSR {\bf 10}, 201 (1946).

\bibitem{Sing} P.~Sindzingre, M.L.~Klein and D.M.~Ceperley, Phys. Rev. Lett. {\bf 63}, 1601 (1981).

\bibitem{Poll} E.L.~Pollock, and D.M.~Ceperley, Phys. Rev. B {\bf 36}, 8343 (1987).

%\bibitem{bernu}D.M.~Ceperley and B.~Bernu, Phys. Rev. Lett. {\bf 93}, 155303 (2004).

\bibitem{wa_first} M.~Boninsegni, N.~Prokof'ev, and B.~Svistunov, Phys. Rev. Lett. {\bf 96}, 070601 (2006); Phys. Rev. E {\bf 74}, 036701 (2006).

\bibitem{wa_app} M.~Boninsegni, N.~Prokof'ev, and B.~Svistunov, Phys. Rev. Lett. {\bf 96}, 105301 (2006); F.~Mezzacapo and M.~Boninsegni, Phys. Rev. Lett. {\bf 97}, 045301 (2006).

\bibitem{Weiss97} C.~Weiss, M.~Wilkens, Optics Express {\bfseries 1}, 272 (1997).

\bibitem{Schneider00} J.~Schneider, H.~Wallis, Eur Phys. J. B {\bfseries 18}, 507--512 (2000).

\bibitem{prl01} A.V.~Filinov, M.~Bonitz, and Yu.E.~Lozovik,
Phys. Rev. Lett. {\bf 86}, 3851 (2001).

\end{thebibliography}
\end{document}